\documentclass[smallcondensed]{svjour3}[14 nov 2007]

\smartqed

\usepackage{graphicx}
\usepackage{epsfig}

\usepackage{dcolumn}

\usepackage{bm}

\usepackage{amsmath}

\usepackage{amssymb}

\begin{document}

\title{Tunable resonators for quantum circuits}

\author{A. Palacios-Laloy \and
        F. Nguyen \and
         F. Mallet \and
          P. Bertet \and 
          D. Vion \and
           D. Esteve}

\institute{Quantronics Group, Service de Physique de l'Etat Condensé (CNRS URA 2464), DRECAM, CEA-Saclay,\\ 91191 Gif-sur-Yvette, France}

\date{Received: date / Accepted: date}

\maketitle

\begin{abstract}

	We have designed, fabricated and measured high-Q $\lambda/2$ coplanar waveguide microwave resonators whose resonance frequency is made tunable with magnetic field by inserting a DC-SQUID array (including $1$ or $7$ SQUIDs) inside. Their tunability range is $30\%$ of the zero field frequency. Their quality factor reaches up to 3$\times10^4$. We present a model based on thermal fluctuations that accounts for the dependance of the quality factor with magnetic field.

	\PACS{74.78.-w, 84.40.Dc, 85.25.Am, 85.25.Dq}

\end{abstract}

\section{Introduction}

On-chip high quality factor superconducting resonators have been extensively studied in the past years due to their potential interest for ultra-high sensitivity multi-pixel detection of radiation in the X-ray, optical and infrared domains \cite{MKIDs,Mazin_thesis}. They consist of a stripline waveguide of well-defined length, coupled to measuring lines through input and output capacitors. The TEM modes they sustain have quality factors defined by the coupling capacitors and reaching in the best cases the $10^6$ range \cite{Mazin_thesis}.

It has also been demonstrated recently \cite{Wallraff04} that superconducting resonators provide very interesting tools for superconducting quantum bit circuits \cite{qubits}. Indeed, a resonator can be used to measure the quantum state of a qubit \cite{Wallraff04,Blais,Lupascu,Siddiqi}. Moreover, another resonator may serve as a quantum bus and mediate a coherent interaction between qubits to which it is coupled. The use of resonators might thus lead to a scalable quantum computer architecture \cite{Blais}. The coupling of two qubits mediated by a coplanar waveguide (CPW) resonator has already been demonstrated \cite{Majer,Sillanpaa}. In experiment \cite{Sillanpaa}, each qubit needs to be tuned in and out of resonance with the resonator for the coupling to be effective. Reference \cite{Wallquist} proposed an alternative solution that consists in tuning the {\it resonator} in and out of resonance with each qubit. Here we report on the measurement of high quality factor resonators whose frequency can be tuned. Measurements similar to ours have been reported by other groups on lumped element \cite{Simmonds} and distributed \cite{Chalmers} resonators.

\section{Tunable resonator with DC SQUID: model \label{sec2}}

\begin{figure}
\centering
   \includegraphics[width=0.7\textwidth,keepaspectratio]{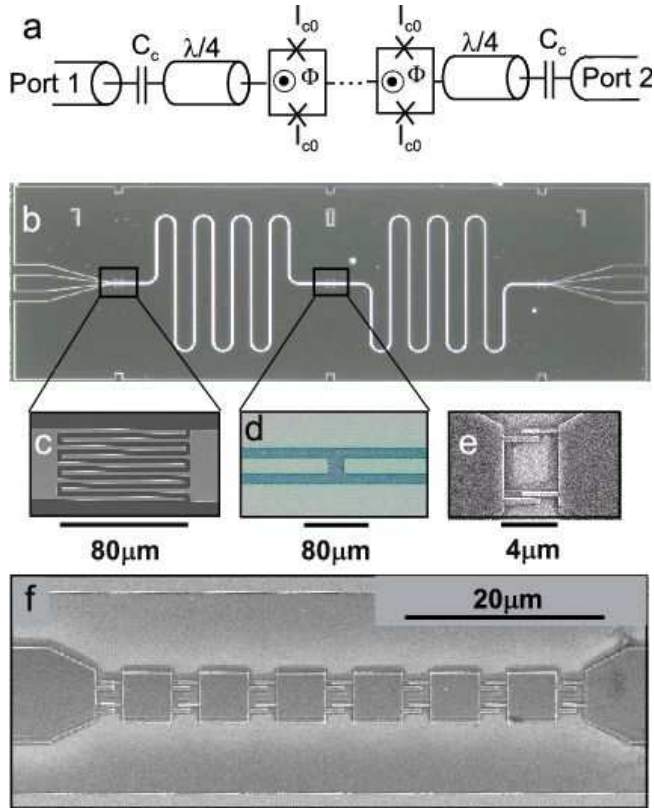}
   \caption{{\bf a:}~Tunable resonator scheme : a DC SQUID array is inserted between two $\lambda/4$ waveguides coupled to a $50$~$\mathrm{\Omega}$ measurement line through input and output capacitors $C_c$. {\bf b:}~Optical micrograph of a CPW niobium resonator. {\bf c:}~Typical coupling capacitor (design value : $C_c=27$fF). {\bf d:}~Gap in the middle of the resonator, before SQUID patterning and deposition. {\bf e:}~Electron micrograph of an aluminum SQUID (sample A), fabricated using electron-beam lithography and double-angle evaporation. {\bf f:}~Electron micrograph of a $7$-SQUID array (sample B).
	\label{fig:figure1}}
\end{figure}

Our tunable resonators consist of $\lambda/2$ coplanar waveguides with an array of $N$ DC-SQUIDs in series inserted in the middle of the central strip (see Fig. \ref{fig:figure1}a). Each DC SQUID is a superconducting loop with self-inductance $L_{l}$ intersected by two nominally identical Josephson junctions of critical current $I_{c0}$ ; the loop is threaded by a magnetic flux $\Phi$. The SQUID array behaves as a lumped non-linear inductance that depends on $\Phi$, which allows to tune the resonance frequency. 

A $\lambda/2$ CPW resonator without any SQUID consists of a transmission line of length $l$, capacitance and inductance per unit length $\mathcal{C}$ and $\mathcal{L}$, and characteristic impedance $Z_0=\sqrt{\mathcal{L}/\mathcal{C}}$. We consider here only the first resonance mode that happens when $l=\lambda/2$ at a frequency $\omega_r = \pi/\sqrt{L C}$, where $L=\mathcal{L} l$ and $C=\mathcal{C} l$ are the total inductance and capacitance of the resonator. The quality factor $Q$ results from the coupling of the resonator to the $R_0=50\,\mathrm{\Omega}$ measurement lines through the input and output capacitors $C_c$ leading to 

\begin{equation}
   Q_{c}=\frac{\pi}{4 Z_0 R_0 C_c^2 \omega_r^2},
\end{equation}

\noindent from internal losses ($Q_{int}$), and from possible inhomogeneous broadening mechanisms ($Q_{inh}$). These combined mechanisms yield

\begin{equation}
Q^{-1} = Q_{c}^{-1} + Q_{int}^{-1} + Q_{inh}^{-1}. 
\end{equation}

\begin{figure}
\centering
   \includegraphics[width=0.7\textwidth,keepaspectratio]{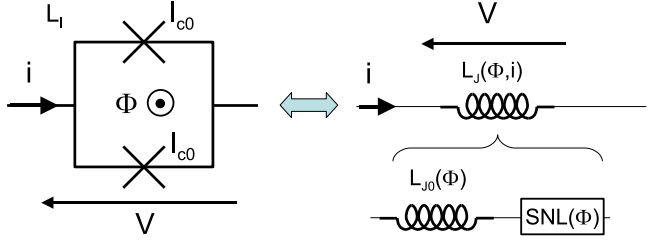}
   \caption{A DC SQUID with two junctions of critical current $I_{c0}$ and loop inductance $L_l$, biased by a magnetic flux $\Phi$ and by a current $i$, is equivalent to a lumped flux-dependent non-linear inductance $L_J(\Phi,i)$ that can be decomposed in an inductance $L_{J0}(\Phi)$ and a non-linear element $\mathrm{SNL}(\Phi)$ in series.  
	\label{fig:figure2}}
\end{figure}

As shown in Fig.~\ref{fig:figure2}, we model a SQUID as a non-linear inductance $L_J(\Phi,i)$ that depends on $\Phi$ and on the current $i$ passing through it, so that the voltage across the SQUID is

\begin{equation}
	V=L_J(\Phi,i) \frac{di}{dt}.
	\label{eq:VI}
\end{equation}

All SQUID properties are periodic in $\Phi$ with a period $\Phi_0  = h/2e$, the superconducting flux quantum. Introducing the reduced flux quantum $\varphi_0 = \Phi_0 / 2\pi$, the SQUID frustration $f = \pi \Phi/\Phi_0$, the effective critical current $I_c(\Phi) = 2 I_{c0} |\cos f|$ of the SQUID at zero loop inductance, and the parameter $\beta = L_l I_{c0} / \varphi_0$, our calculation of $L_J(\Phi,i)$ to first order in $\beta$ and to second order in $i/I_c(\Phi)$ yields for $f \in \left] -\pi/2,\pi/2\right[$

\begin{equation}
	L_J(\Phi,i) = L_{J0}(\Phi) + A(\Phi) i^2, 
	\label{eq:LJ}
\end{equation}

\noindent with

\begin{eqnarray}
	L_{J0}(\Phi) & = & \frac{\varphi_0}{I_c(\Phi)} \left(1 + \beta \frac{\cos 2f}{2 \cos f} \right),  \\
  A(\Phi)	& = & \frac{\varphi_0}{2 I_c^3(\Phi)}.
\end{eqnarray}

Equation \ref{eq:LJ} shows that the SQUID can be modelled as the series combination of a lumped inductance $L_{J0}(\Phi)$ and of a non-linear device SNL($\Phi$) \cite{Vladimir} (see Fig.~\ref{fig:figure2}). 

In the linear regime $i \ll I_c(\Phi)$ corresponding to low intra-cavity powers, one can neglect the non-linear term in Eq.~\ref{eq:LJ}. The N-SQUID array then simply behaves as a lumped inductance $N L_{J0}(\Phi)$. The device works in that case as a tunable harmonic oscillator. Introducing the ratio $\varepsilon(\Phi)=L_{J0}(\Phi)/L$ between the total effective inductance of the SQUID and the resonator inductance, the frequency and quality factor are

\begin{eqnarray}
	\omega_0(\Phi) & = & \omega_r \frac{1}{1+N \varepsilon(\Phi)},  \label{eq:omega_vs_flux} \\
	Q_{ext}(\Phi) & = & Q_c \left[ 1 +  4 N \varepsilon(\Phi) \right].
	\label{eq:Q_vs_flux}
\end{eqnarray}

At larger peak current in the resonator $i \lesssim I_c(\Phi)$, the non-linear element SNL($\Phi$) has to be taken into account. The equation of motion of the oscillator acquires a cubic term, similar to that of a Duffing oscillator \cite{Landau}. This leads to a small additional shift of the resonance frequency $\delta \omega_0 (E)$ proportional to the total electromagnetic energy $E$ stored in the resonator. Retaining first order terms in $\varepsilon(\Phi)$, we find

\begin{eqnarray}
	\frac{\delta \omega_0 (\Phi,E)}{\omega_0(\Phi)} & = & - N \left\{\frac{2 \omega_0(\Phi)}{\pi R_0 [1 + 2 N \varepsilon(\Phi)]} \right\}^2  \frac{\varphi_0}{8 I_c^3(\Phi)} E.
	\label{eq:anhar_shift}
\end{eqnarray}

As shown by Eq.~\ref{eq:omega_vs_flux}, a resonator including an array of $N$ SQUIDs of critical current $N I_{c0}$ has approximately the same resonant frequency and same tunability range as a resonator including one SQUID of critical current $I_{c0}$. However, an interesting advantage of using an array is to obtain a linear regime that extends to larger currents, allowing measurements at larger powers and therefore higher signal-to-noise ratios.

\section{Sample fabrication}

The design and fabrication of our resonators closely followed Ref. \cite{Frunzio}. The coupling capacitors were simulated using an electromagnetic solver. Test niobium resonators without any SQUIDs were first fabricated. They were patterned using optical lithography on a $200\,$nm thick niobium film sputtered on a high-resistivity ($>1000~\mathrm{\Omega}~\mathrm{cm}$) oxidized 2-inch silicon wafer. The niobium was etched away using either dry or wet etching. Dry etching was done in a plasma of pure SF$_6$ at a pressure of $0.3\,$mbar and at a power such that the self-bias voltage was $30\,$V and the etching rate $1.3\,$nm/s. We observed that adding oxygen to the plasma gave consistently lower quality factors. Wet etching was done in a solution of HF, H$_2$O, and FeCl$_3$ having an etching rate of approximately $1\,$nm/s at room-temperature. A typical resonator and its coupling capacitor are shown in panels b and c of Fig.~\ref{fig:figure1}. Its $3.2\,$cm length yields a resonance frequency around $1.8\,$GHz. 

In addition to these test structures, some resonators had a gap in the middle (see Fig.~\ref{fig:figure1}d) used in a later step to fabricate a SQUID array by e-beam lithography and double-angle aluminum deposition (see panels e and f in Fig.~\ref{fig:figure1}). Before depositing the aluminum, the niobium surface was cleaned by argon ion-milling (dose $\lesssim 10^{18}$ neutralized $500\,$eV ions per square centimeter). The Nb/Al contact resistance was found to be in the ohm range, yielding tunnel junctions of negligible inductance compared to that of the SQUID.

\section{Experimental setup}

\begin{figure}
\centering
   \includegraphics[width=0.9\textwidth,keepaspectratio]{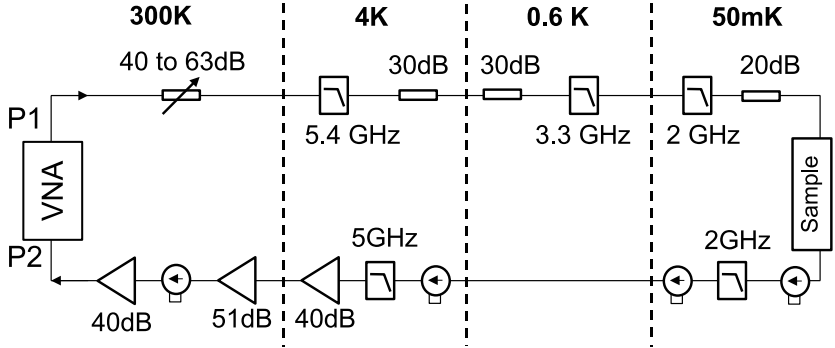}
	\caption{Experimental setup. The sample is thermally anchored at the mixing chamber of a dilution refrigerator with temperature $40 - 60\,$mK. It is connected to a vector network analyzer (VNA) at room-temperature that measures the amplitude and phase of the S$_{21}$ coefficient. The input line (top) is strongly attenuated ($120$ to $160\,$dB in total) with cold attenuators to protect the sample from external and thermal noise, and filtered above $2\,$GHz. The output line (bottom) includes a cryogenic amplifier with a $3\,$K noise temperature and $3$ cryogenic isolators. \label{fig:figure3}}
\end{figure}

The chips were glued on a TMM$10$ printed-circuit board (PCB). The input and output port of the resonator were wire-bonded to coplanar waveguides on the PCB, connected to coaxial cables via mini-SMP microwave launchers. The PCB was mounted in a copper box. The $S_{21}$ coefficient (amplitude and phase) of the scattering matrix was measured as a function of frequency using a vector network analyzer. Test resonators were measured in a pumped $^4$He cryostat reaching temperatures of $1.3\,$K, with typical input power of $-50\,$dBm and using room-temperature amplifiers. We measured internal quality factors up to 2$\times10^5$ with both etching methods. 

The tunable resonators were measured in a dilution refrigerator operated at $40-60\,$~mK, using the microwave setup shown in Fig.~\ref{fig:figure3}. The input line includes room-temperature and cold attenuators. The output line includes $3$ cryogenic isolators, a cryogenic amplifier (from Berkshire) operated at $4\,$K with a noise temperature of $3\,$K, and additional room-temperature amplifiers. The attenuators and isolators protect the sample from external and thermal noise. This setup allowed to measure the sample with intra-cavity energies as small as a few photons in order to operate in the linear regime, corresponding to typical input powers of $-140\,$dBm at the sample level.

\begin{table}
	\centering
		\begin{tabular}{|c|c|c|c|c|c|c|c|}
			\cline{2-8}
			  \multicolumn{1}{c|}{\rule{0pt}{2.5ex}} & \multicolumn{4}{c|}{Design} & \multicolumn{3}{c|}{Measurements}  \\
			\cline{2-8}
			  \multicolumn{1}{c|}{\rule{0pt}{2.5ex}} & $C_c$ & $Q_c$ & $L_l$ & $N$ & $I_{c0}$ & $\omega_r/2\pi$ & $Q(\Phi=0)$ \\
			  \hline
			  \rule{0pt}{2.5ex} Test & $2\,$fF & 6$\times10^5$ & & $0$ & & $1.906\,$GHz & 2$\times10^5$ \\
			  \rule{0pt}{2.5ex} Sample A & $27\,$fF & 3.4$\times10^3$ & $40 \pm 10\,$ pH & $1$ & $330\,$nA & $1.805\,$GHz & 3.5$\times10^3$ \\
			  \rule{0pt}{2.5ex} Sample B  & $2\,$fF & 6$\times10^5$ & $20 \pm 10\,$ pH & $7$ & $2.2 \mu$A & $1.85\,$GHz & 3$\times10^4$ \\
			  \hline
		\end{tabular}
		\caption{Summary of sample parameters. See text for definitions.
		\label{table:parameters}}
\end{table}

\section{Experimental results}

\begin{figure}
\centering
   \includegraphics[width=1\textwidth,keepaspectratio]{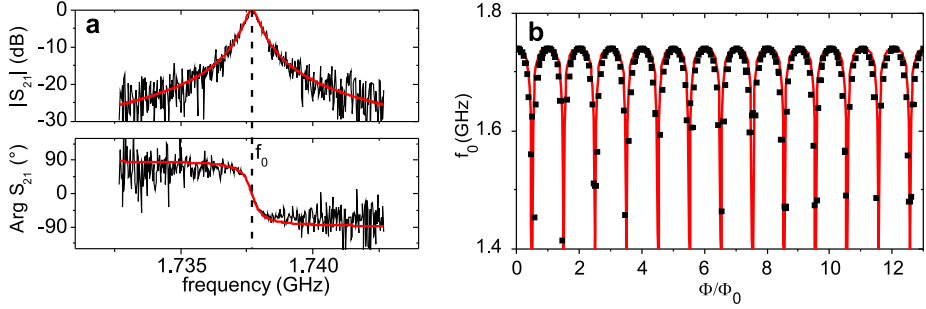}
	\caption{(color online) {\bf a:}~Measured (thin line) amplitude (top) and phase (bottom) transmission of sample A for $\Phi=0$ and fit (bold line) yielding a quality factor $Q=3300$. {\bf b:}~Measured resonance frequency of sample A (squares) as a function of applied magnetic flux and corresponding fit (full line) according to Eq.~\ref{eq:omega_vs_flux}.
	\label{fig:figure4}}
\end{figure}

Two tunable resonators were measured: sample $A$ has only one SQUID (see Fig.~\ref{fig:figure1}e) and large coupling capacitors ($27\,$fF) so that its total quality factor is determined by $Q_c=$~3.4~$\times10^3$. Sample $B$ has an array of $7$ SQUIDs in series (see Fig.~\ref{fig:figure1}f) and smaller coupling capacitors ($2\,$fF) so that its quality factor is likely to be dominated by internal losses or inhomogeneous broadening. Relevant sample parameters are listed in table \ref{table:parameters}.

A typical resonance curve, obtained with sample $A$ at $\Phi=0$ for an input power of $-143\,$dBm corresponding to a mean photon number in the cavity $\overline{n} \approx 1.2$, is shown in Fig.~\ref{fig:figure4}. The $|$S$_{21}|$ curve was normalized to the maximum measured value. By fitting both the amplitude and the phase response of the resonator, we extract the resonance frequency and the quality factor $Q$. When the flux through the SQUID is varied, the resonance frequency shifts periodically as shown in Fig.~\ref{fig:figure4}b, as expected.

\begin{figure}
\centering
   \includegraphics[width=0.9\textwidth,keepaspectratio]{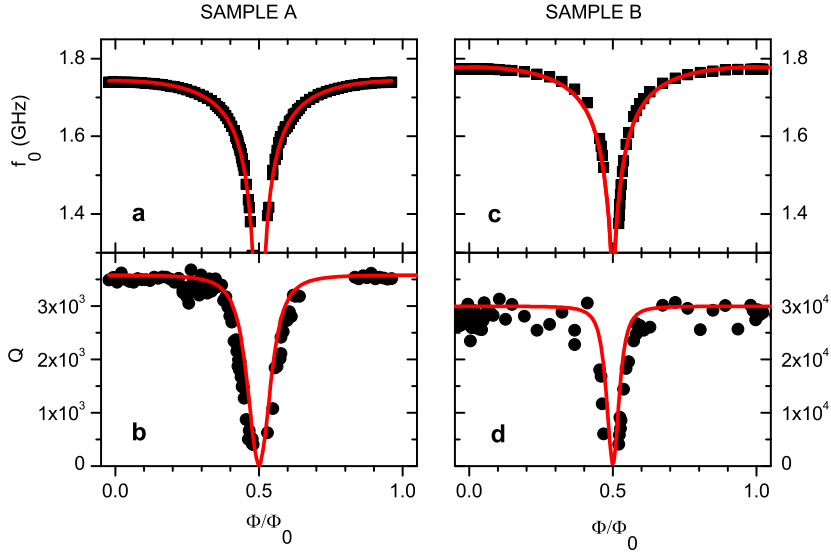}
	\caption{(color online) {\bf a} and {\bf c:} Measured resonance frequency $f_0$ as a function of $\Phi/\Phi_0$ (squares) for samples A and B, respectively, and fit according to Eq.~\ref{eq:omega_vs_flux} (solid line). {\bf b} and {\bf d:} Measured quality factor $Q$ (disks) as a function of $\Phi/\Phi_0$. The solid line is calculated according to the model (see text) for a temperature $T=60\,$mK.
	\label{fig:figure5}}
\end{figure}

The resonance frequency $f_0(\Phi)$ and quality factor $Q(\Phi)$ are shown for both samples in Fig.~\ref{fig:figure5} over one flux period. The $f_0(\Phi)$ curves in panels a and c are fitted with Eq.~\ref{eq:omega_vs_flux}. The agreement is good over the whole frequency range, which extends from $1.3$ to $1.75\,$GHz, yielding a tunability range of $30\%$. The small discrepancy observed for sample $B$ might be due to a dispersion in the various SQUID loop areas that is not taken into account in our model. The parameters obtained by this procedure for both samples are shown in table \ref{table:parameters}; they are in good agreement with design values and test-structure measurements.

The $Q(\Phi)$ dependance for both samples is shown in panels b and d of Fig.~\ref{fig:figure5}. Both samples show a similar behaviour: the quality factor depends weakly on $\Phi$ when the flux is close to an integer number of flux quanta, whereas it shows a pronounced dip around $\Phi_0/2$. 

The largest quality factors are 3.5$\times10^3$ for sample $A$ and 3$\times10^4$ for sample $B$. This difference is due to the different coupling capacitors. For sample $A$, the maximum quality factor is the same as measured on test resonators with similar capacitors and corresponds to the expected $Q_c$ for $C_c=27\,$fF. Therefore sample $A$ quality factor is limited by the coupling to the $50\,\mathrm{\Omega}$ lines around integer values of $\Phi_0$. The situation is different for sample $B$: the measured value is one order of magnitude lower than both the quality factor $Q_c=$6$\times 10^5$ expected for $C_c=2\,$fF and the measured $Q$ of test resonators with the same capacitors (see table \ref{table:parameters}). This unexplained broadening of the resonance in presence of a SQUID array might be due either to the presence of low-frequency noise in the sample, or to a dissipation source specifically associated with the SQUIDs. We note that flux-noise is not plausible since our measurements show no clear correlation with the sensitivity of the resonator to flux-noise. However, critical-current noise could produce such effect. Another possibility could be dielectric losses in the tunnel barriers. 

We now turn to the discussion of the dip in $Q(\Phi)$ observed around $\Phi_0/2$. We attribute it to thermal noise. Indeed, as discussed in section \ref{sec2}, the resonance frequency depends on the energy stored in the resonator. At thermal equilibrium, fluctuations in the photon number translate into a fluctuation of the resonance frequency and cause an inhomogeneous broadening. At temperature $T$, the resonator stores an average energy given by Planck's formula $\overline{E}=\hbar \omega_0(\Phi) \overline{n}$, $\overline{n}=1/\{\exp[\hbar \omega_0(\Phi)/k T] - 1\}$ being the average photon number. The photon number and energy fluctuations are $\overline{n^2 - \overline{n}^2} = \overline{n} (\overline{n} + 1)$ and

\begin{equation}
   \sqrt{\overline{\delta E^2}} = \sqrt{\overline{E}^2 + \hbar \omega_0(\Phi) \overline{E}}.
\end{equation}

The characteristic time of these energy fluctuations being given by the cavity damping time $Q/\omega_0$ with $Q \gg 1$, a quasi-static analysis is valid and leads to an inhomogeneous broadening $\delta \omega_{inh} = \left|d \omega_0 /d E\right|\sqrt{\overline{\delta E^2}}$. Using Eq.~\ref{eq:anhar_shift}, we get 

\begin{equation}
			Q_{inh}^{-1}(\Phi) = \frac{\delta \omega_{inh}(\Phi)}{\omega_0(\Phi)} = 
       N \left\{\frac{2 \omega_0(\Phi)}{\pi R_0 [1 + 2 N \varepsilon(\Phi)]} \right\}^2 \frac{\varphi_0}{8 I_c^3(\Phi)}\sqrt{\overline{\delta E^2}}.
       \label{eq:deltaomega_inh}
\end{equation}

The resulting quality factor is $Q^{-1}=Q_{inh}^{-1}+Q_{ext}^{-1}$, which is plotted as full curves in panels b and d of Fig.~\ref{fig:figure5}, for $T=60\,$mK. The agreement is good, although Eq.~\ref{eq:deltaomega_inh} results from a first-order expansion that is no longer valid in the close vicinity of $\Phi_0/2$. We have also observed that $Q$ values significantly degrade around $\Phi_0/2$ when the samples are heated, while remaining unchanged around integer numbers of $\Phi_0$. These observations suggest that thermal noise is the dominant contribution to the drop of $Q$. Note that our model does not take into account flux-noise, which evidently contributes to $Q_{inh}$ and could account for the residual discrepancy between experimental data and theoretical curves in panels b and d of Fig.~\ref{fig:figure5}.

\section{Conclusion}

We have designed and measured SQUID-based stripline resonators that can be tuned between $1.3\,$GHz and $1.75\,$GHz, with a maximum $Q=$3$\times10^4$ limited by an unknown mechanism. The quality factor degrades due to thermal noise around $\Phi_0/2$. This limitation would be actually lifted with higher frequency resonators matching typical Josephson qubit frequencies. Their tunability range at high $Q$ would then be wide enough to couple a large number of qubits. 

\begin{acknowledgement}
This work has been supported by the European project EuroSQIP. We acknowledge technical support from P. S\'enat, P.F. Orfila and J.C. Tack, and fruitful discussions within the Quantronics group and with A. Lupascu, A. Wallraff, M. Devoret, and P. Delsing.
\end{acknowledgement}

\end{document}